%% file: main_latex.tex
\pdfoutput=1

\documentclass[11pt]{article}

\usepackage[final]{acl}

\usepackage{times}
\usepackage{latexsym}

\usepackage[T1]{fontenc}

\usepackage[utf8]{inputenc}

\usepackage{microtype}

\usepackage{inconsolata}

\usepackage{graphicx}

\usepackage[normalem]{ulem}
\usepackage{multirow}
\usepackage{booktabs}
\usepackage{caption} 
\usepackage{arydshln}
\usepackage{amssymb}
\usepackage{makecell}
\usepackage{array}
\usepackage{float}
\usepackage{url}
\usepackage{algorithm}
\usepackage{algorithmic}
\usepackage{amsmath}
\usepackage{enumitem}
\usepackage{eqparbox}

\usepackage{pifont}

\newcommand{\cmark}{\ding{51}}%
\newcommand{\xmark}{\ding{55}}%
\newcolumntype{C}[1]{>{\centering\arraybackslash}m{#1}}

\input{macro}

\title{Hierarchical Retrieval with Evidence Curation for Open-Domain Financial Question Answering on Standardized Documents}

\author{
  Jaeyoung Choe, Jihoon Kim, Woohwan Jung \\
  Department of Applied Artificial Intelligence, Hanyang University \\
  \texttt{\{cjy9100, skygl, whjung\}@hanyang.ac.kr}
  }

\begin{document}
\maketitle
\begin{abstract}
Retrieval-augmented generation (RAG) based large language models (LLMs) are widely used in finance for their excellent performance on knowledge-intensive tasks.
However, standardized documents (e.g., SEC filing) share similar formats such as repetitive boilerplate texts, and similar table structures.
This similarity forces traditional RAG methods to misidentify near-duplicate text, leading to duplicate retrieval that undermines accuracy and completeness.
To address these issues, we propose the \textbf{Hi}erarchical \textbf{R}etrieval with \textbf{E}vidence \textbf{C}uration (HiREC) framework. 
Our approach first performs hierarchical retrieval to reduce confusion among similar texts. 
It first retrieve related documents and then selects the most relevant passages from the documents. 
The evidence curation process removes irrelevant passages. When necessary, it automatically generates complementary queries to collect missing information.
To evaluate our approach, we construct and release a \textbf{L}arge-scale \textbf{O}pen-domain \textbf{Fin}ancial (\benchmark) question answering benchmark that includes 145,897 SEC documents and 1,595 question-answer pairs. 
Our code and data are available at \url{https://github.com/deep-over/LOFin-bench-HiREC}.

\end{abstract}

\input{latex/1_introduction}

\input{latex/2_related_works}

\input{latex/3_dataset}

\input{latex/4_framework}

\input{latex/5_experiments}

\input{latex/6_conclusion_and_others}

\input{latex/8_limitation}

\bibliography{citation}

\input{latex/7_appendix}

\end{document}

%% file: macro.tex
\newcommand{\ours}{HiREC~}

\newcommand{\benchmark}{LOFin}
\newcommand{\benchmarkold}{LOFin-1.4k~}
\newcommand{\benchmarknew}{LOFin-1.6k~}
\newcommand{\benchmarkfullname}{Large-scale Open-domain Financial QA}

\renewcommand{\algorithmiccomment}[1]{\hfill\eqparbox{COMMENT}{// #1}}

%% file: latex/1_introduction.tex
\section{Introduction}

Retrieval-augmented generation (RAG) ~\cite{lewis2020retrieval} with large language models (LLMs) have significantly improved performance in knowledge-intensive tasks.
Due to its ability to improve both factual accuracy and timeliness, extensive research ~\cite{yepes2024financial,sarmah2024hybridrag} has investigated applying RAG in searching financial information where accurate and up-to-date information is crucial for decision-making.

\begin{figure}[!tbp]
    \centering
    \includegraphics[width=\linewidth]{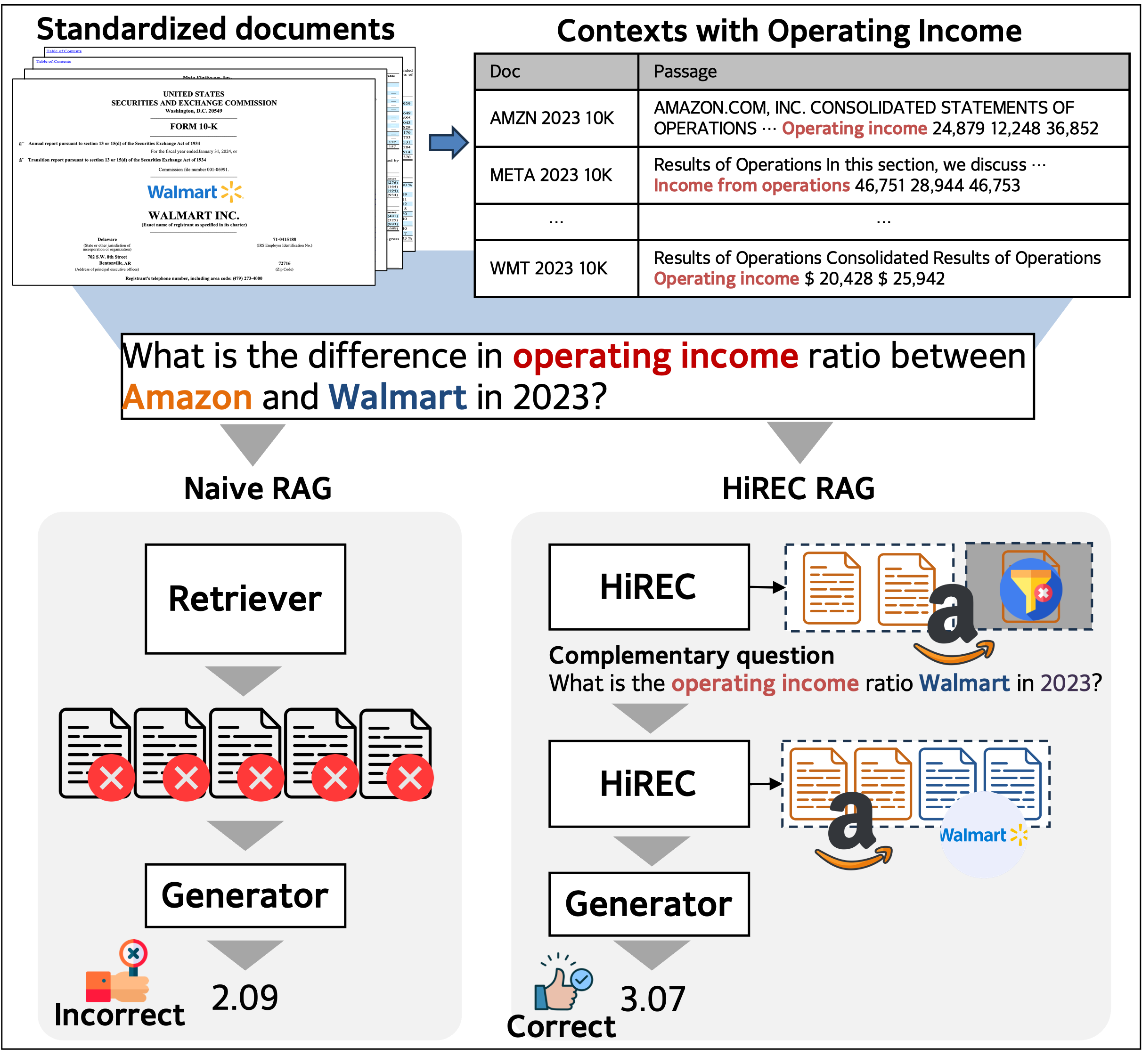}
    \vspace{-0.5cm}
    \caption{Comparison of a naive RAG approach and \ours.}
    \vspace{-0.5cm}
    \label{fig:enter-label}
\end{figure}

Financial documents, such as annual reports (10-K) in SEC filings, are highly structured and follow standardized templates across companies and periods, often containing similar tables and repetitive narratives. 
As shown in \autoref{fig:enter-label}, 2023 10-K reports from Amazon, Meta, and Walmart exhibit nearly identical table structures with similar titles and indicators, differing mainly in numerical values.
Consequently, when asked, \textit{What is the difference in operating income ratio between Amazon and Walmart in 2023?}, a conventional RAG system may struggle to distinguish among these similar passages, retrieving irrelevant or redundant information that leads to inaccurate answers.

To tackle these challenges from the standardized format of financial documents, we propose the HiREC (Hierarchical Retrieval and Evidence Curation) framework. HiREC consists of two main components: hierarchical retrieval and evidence curation. The hierarchical retrieval first retrieves related documents and then selects the most relevant passages from the documents, thereby reducing confusion from near-duplicate text. 
As illustrated in \autoref{fig:enter-label}, narrowing the candidate set to documents from Amazon and Walmart enables the system to focus on highly relevant passages. 
However, the prevalence of comparative questions in financial QA often leads to incomplete retrieval of essential evidence. To address this, the evidence curation stage filters out irrelevant passages and generates complementary queries when necessary. 
For example if only Amazon operating income is retrieved then a complementary query is generated to fetch Walmart operating income so that all necessary evidence is gathered for an accurate answer.

To evaluate our approach, we assess QA performance in an open-domain setting. Existing financial question-answering benchmarks~\cite{islam2023financebench, lai2024sec} rely on small-scale corpora that include at most about 1,300 documents and very limited test sets, which do not reflect realistic financial scenarios. We propose \benchmark~(\benchmarkfullname), a comprehensive financial question-answering benchmark built on a large-scale corpus containing approximately 145,000 SEC filings from companies in the S\&P 500. \benchmark~includes 1,595 open-domain question-answering test instances and addresses challenges in standardized document retrieval, such as near-duplicate tables and repetitive narratives, that are not evident in smaller datasets. In addition, we release the entire benchmark as open-source to support future research in the field.

Experimental results indicate that HiREC improves performance by at least 13\% compared to existing RAG methods. 
In addition, our framework consistently outperforms commercial llms with web search engines such as SearchGPT~\cite{searchGPT} and Perplexity~\cite{Perplexity2023}.

Our contributions are threefold:
\begin{itemize}[noitemsep]
\vspace{-0.15cm}
    \item We propose a hierarchical retrieval approach that retrieves related documents and identifies the most pertinent passages, thereby reducing confusion caused by near-duplicate content in standardized financial documents.
    \item We introduce an evidence curation process that filters out irrelevant passages and generates complementary queries when necessary, effectively supplementing missing information for accurate financial QA.
    \item We present \benchmark, a large-scale, realistic financial QA benchmark that exposes challenges in standardized document retrieval, and we release it as open-source.
\end{itemize}

%% file: latex/2_related_works.tex
\section{Related Works}

\paragraph{Financial QA and RAG}
Financial QA has advanced significantly, with recent benchmarks emphasizing numeric reasoning and table understanding. TAT-QA~\cite{zhu2021tat} and FinQA~\cite{chen2021finqa} provide single-page tabular contexts, while DocFinQA~\cite{reddy2024docfinqa} and DocMath-Eval~\cite{zhao-etal-2024-docmath} extend this to multi-page settings. However, these benchmarks remain closed-domain, limiting their applicability for RAG systems. Open-domain benchmarks have also been proposed. For instance, consider Financebench~\cite{islam2023financebench} and SEC-QA~\cite{lai2024sec}. However, these datasets are built on small-scale document collections and suffer from limited test set sizes or the lack of publicly available fixed test sets.

With the emergence of financial QA, research on financial RAG has also been progressing.
There are graph-based methods~\cite{barry2025graphrag, sarmah2024hybridrag} tailored to the financial domain as well as hybrid approaches~\cite{wang2024financerag}. In addition, studies on financial document chunking offer valuable insights~\cite{yepes2024financial}.

\paragraph{Retrieval augmented generation}
RAG has evolved in diverse ways~\cite{gao2023retrieval, zhang2025survey}.
Hierarchical retrieval was proposed for cases where sections are clearly demarcated, typically employing a two-step document-passage process~\cite{arivazhagan-etal-2023-hybrid, chen2024hiqa}. 
The difference from previous studies is that while they segment documents into sections, we segment at the level of individual documents.
There are also studies that utilize filtering to enhance the quality of the contexts input into RAG~\cite{zhuang-etal-2024-efficientrag, wang2024rear}. Unlike previous approaches that required training, we manage quality using only an LLM.
Iterative retrieval is typically proposed for multi-hop QA. 
A standard iterative method uses the context retrieved in the first step as part of the query for subsequent iterations~\cite{trivedi2022interleaving, shao2023enhancing}. 
Self-RAG~\cite{asai2023self} also conducts iterative retrieval that includes the generation process. However, our approach does not utilize previously retrieved context because we specifically need to discover missing information.

%% file: latex/3_dataset.tex
\section{Large-scale Open-domain Financial QA}

\label{sec:benchmark}
In this section, we present \benchmark, a benchmark that overcomes the limitations of current financial QA datasets by expanding the retrieval corpus and increasing open-domain QA pairs (see \autoref{table:benchmark_comparison}).

\subsection{Large-scale Document Collection}
To reflect a real-world scenario where retrieval and QA must be performed over a large volume of documents, we collected a comprehensive set of SEC filings. Specifically, we gathered 10-K, 10-Q, and 8-K filings from the SEC EDGAR\footnote{\url{https://www.sec.gov/search-filings}} system, covering S\&P 500 companies from October 2001 to April 2025.

We converted the HTML documents to PDF using the wkhtmltopdf\footnote{\url{https://wkhtmltopdf.org/}} library. Following the approach of \citealp{islam2023financebench}, we used the PyMuPDF library\footnote{\url{https://pymupdf.readthedocs.io/}} to extract text at the page level from these PDFs, excluding reports that lacked proper page separation. The final corpus consists of 145,897 reports from 516 companies.

\subsection{Open-domain QA Pair Construction} \label{benchmark_question_construction}

In this section, we detail our process for constructing open-domain QA pairs by leveraging three established financial QA benchmarks: FinQA~\cite{chen2021finqa}, Financebench~\cite{islam2023financebench}, and SEC-QA~\cite{lai2024sec}.

We begin by converting the closed-setting questions from FinQA into an open-domain format. First, we exclude any test questions for which evidence documents were not collected due to page separation or collection period issues (35 out of 1147). Next, we transform the remaining questions using GPT-4o by appending relevant period and company information. For example, the question \textit{what are the total operating expenses for 2016?} is transformed into \textit{Could you provide the total operating expenses reported by Lockheed Martin for the year 2016?}, with Lockheed Martin explicitly added to enhance context. The conversion prompt used is provided in Appendix~\ref{appendix:bench_finqa_prompt}.

Subsequently, we identify candidate evidence pages using a two-step process. 
First, we compute BM25 similarity scores between the FinQA gold table context and the content of each page in the candidate document, explicitly considering the distinct numerical values in the table. 
Then, we employ an NLI model\footnote{\url{https://huggingface.co/sentence-transformers/nli-mpnet-base-v2}} to measure the semantic similarity between the FinQA gold context (excluding table content) and each page. 
If both methods select the same top candidate page we accept it as evidence and verify its correctness. Otherwise we manually annotate the correct page to ensure accurate evidence labeling (see Appendix~\ref{appendix:bench_finqa_anno} for further details).

\begin{table}[!t]
  \centering
  \input{tables/dataset_statistic_comparison}
  \label{table:benchmark_comparison}
\end{table}


Financebench is designed for an open-domain setting and its questions are adopted without modification. 
In contrast both FinQA and Financebench mainly include single-document questions which limits their ability to evaluate multi-document retrieval and reasoning. 
To address this limitation, we adopt multi-document question templates from SEC-QA. 
We select the four that are designed for multi-document scenarios. 
We then manually craft the associated questions and annotate the answers along with the corresponding evidence pages. 
This process enhances our open-domain QA pairs with challenges that require multi-document and multi-hop reasoning.

Following the ACL ARR review, we finalized the \benchmark~benchmark with a total of 1,595 QA pairs (initially 1,389), reflecting the addition of 205 newly created QA pairs based on recent SEC filings.
We refer to the initial version as \benchmarkold and the expanded version as \benchmarknew to clearly distinguish between the two.
These newly added questions follow the same annotation protocol and are designed to promote more complex reasoning across multiple documents.
For details of the expanded \benchmarknew benchmark, see Appendix~\ref{appendix:expanded_lofin}; for the SEC-QA templates used in its construction, refer to Appendix~\ref{appendix:bench_secqa}.

%% file: tables/dataset_statistic_comparison.tex

\Large
\renewcommand{\arraystretch}{1.2}
\resizebox{\columnwidth}{!}{
\begin{tabular}{lccc r}
\toprule
\textbf{Benchmark} & \textbf{Open} & \textbf{Multi-Doc} & \textbf{\# QAs} & \textbf{\# Docs} \\
\midrule
TAT-QA       &        &        & 1,669  & -      \\
FinQA        &        &        & 1,147  & -      \\
DocFinQA     &        &        & 922    & -      \\
Financebench & $\checkmark$ &        & 150    & 368    \\
SEC-QA       & $\checkmark$ & $\checkmark$ & N/A    & 1,315  \\
\hline
\benchmark         & $\checkmark$ & $\checkmark$ & 1,595  & 145,897 \\
\bottomrule
\end{tabular}
}
\caption{Comparison of financial QA benchmarks. “\# QAs” is the test set size, and “\# Docs” is the number of documents in the retrieval corpus. SEC QA does not have a fixed QA count, as it provides a question generation framework.}
\vspace{-0.5cm}
\label{tab:qa_datasets_comparison}

%% file: latex/4_framework.tex
\begin{figure*}[!]
    \centering
    \includegraphics[width=\textwidth]{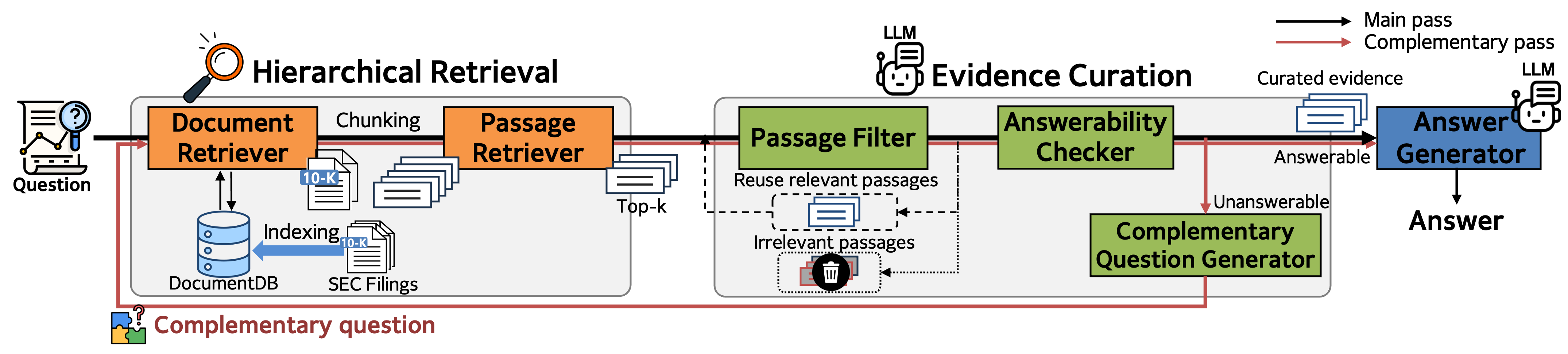}
    \caption{Overview of hierarchical retrieval with evidence curation framework}
    \vspace{-0.5cm}
    \label{fig:figure1}
\end{figure*}

\section{Hierarchical Retrieval with Evidence Curation (HiREC) Framework}

In this section, we introduce the HiREC framework.
\autoref{fig:figure1} shows the overall framework and process.
The framework comprises two main components: hierarchical retrieval and evidence curation. 
During the hierarchical retrieval stage a hierarchical approach retrieves passages \( \mathcal{P}_r \) that are relevant to the question \( q \).
In the evidence curation process the retrieved passages are filtered to retain only those directly pertinent and then evaluated to determine whether they provide sufficient information to answer the question.
If the information is insufficient a complementary question \( q_{c} \) is generated to reinitiate the retrieval process (complementary pass).
Otherwise if the evidence is sufficient the filtered passage set \( \mathcal{P}_f \) is forwarded to the Answer Generator to produce the final answer (main pass).
When the maximum iteration \(i_{\text{max}}\) is reached evidence curation halts and the passages retrieved using \(q_{\text{c}}\) are merged with the previously filtered passages to generate the answer. 
The pseudocode of HiREC is described in \autoref{alg:overall}.
For the LLM-based components of our framework, we use instruction-style prompts tailored to each module's objective. \autoref{appendix:prompt_list} provides the full prompt templates and design process.

\subsection{Hierarchical Retrieval}

\label{sec:hr}
Standardized documents use uniform templates with repetitive structures and similar content, which makes retrieving distinct and relevant passages challenging. We address this issue using a hierarchical approach. First we retrieve documents relevant to the question (\ref{sec:dr}) to narrow the search space, and then we select pertinent passages within those documents (\ref{sec:pr}).

\subsubsection{Document Retriever} \label{document_retriever}
\paragraph{Document indexing.}
\begin{algorithm}[!htb]
\captionsetup{font=small}
\caption{HiREC framework}
\label{alg:overall}
\footnotesize
\begin{algorithmic}[1]
\REQUIRE A question $q$, a corpus $\mathcal{D}$, maximum number of iterations $i_{max}$
\STATE $\mathcal{P}_f \gets \emptyset$ // Initialization
\STATE $\mathcal{P}_r \gets$ \textsc{HierarchicalRetrieval}$(q, \mathcal{D})$  \algorithmiccomment{Sec~\ref{sec:hr}}
\FOR{$i = 1 ...  {i}_{max}$}
    \STATE ($\mathcal{P}_f$, $q_{c}$, $y$) $\gets$ \textsc{EvidenceCuration}$(q, \mathcal{P}_r)$ \algorithmiccomment{Sec~\ref{sec:evidencecuration}}\\ 
    \IF{$y$ is \texttt{Answerable}}
        \RETURN \textsc{AnswerGeneration}$(q, \mathcal{P}_f)$ \algorithmiccomment{Sec~\ref{sec:ag}}
    \ENDIF
    \STATE $\mathcal{P}_r \!\gets\!$ \textsc{HierarchicalRetrieval}$(q_{c}, \mathcal{D})$ \algorithmiccomment{Sec~\ref{sec:hr}}
    \STATE $\mathcal{P}_r \gets \mathcal{P}_r \cup \mathcal{P}_f$
\ENDFOR
\RETURN \textsc{AnswerGeneration}$(q, \mathcal{P}_r)$ \algorithmiccomment{Sec~\ref{sec:ag}}
\end{algorithmic}
\end{algorithm}
\setlength{\textfloatsep}{6pt}

Documents contain extensive context, and their standardized format makes it difficult to capture all important details in a single vector. 
To address this, we extract and index key distinguishing information when retrieving documents. 
In financial reports, the cover page provides essential details such as the company name, report type, and fiscal period. 
For each document $d \in \mathcal{D}$, we generate a cover page summary $d'$ using an LLM (the prompt is detailed in Appendix~\ref{appendix:summarization_prompt}), precompute its embedding with a bi-encoder~\cite{wang2024multilingual}, and index the resulting vector as $\mathbf{v}_d = E^{D}(d')$ in the document store.


\paragraph{Document retrieval process.}
\label{sec:dr}
Our document retrieval process consists of three stages. First, given a question \(q\), we use an LLM to convert it into a refined query \(q'\). This conversion reduces issues caused by extraneous financial terms (e.g., tickers or service names like Google or Facebook) that may hinder effective retrieval. (Details of the transformation prompt are provided in Appendix~\ref{appendix:query_transform_prompt}.) Second, we perform dense retrieval. We compute the vector representation of \(q'\) using the same bi-encoder employed during document indexing, yielding \(\mathbf{v}_{q'} = E^{D}(q')\). The relevance between \(q'\) and each document \(d\) is then determined via \(s^D_{q',d} = \mathbf{v}_{q'}^\top \mathbf{v}_d\)~\cite{karpukhin2020dense}; based on these scores, we retrieve \(k'_D\) candidate documents \(\mathcal{D}_{\text{cand}}\). Finally, we rerank these candidates using a cross-encoder~\cite{he2021debertav3} by computing \(\text{CrossEncoder}^D(q', d)\) and select the top \(k_D\) documents \(\mathcal{D}_r\). This multi-stage process ultimately produces the final set of documents \(\mathcal{D}_r\) that are most relevant to the original question \(q\).

\subsubsection{Passage Retriever}
\label{sec:pr}

Within the final set of retrieved documents \(\mathcal{D}_r\), the passage retriever evaluates each passage \(p\) by computing a score using a cross-encoder, \(\text{CrossEncoder}^P(q, p)\). It then selects the top \(k_P\) passages to form the final set \(\mathcal{P}_r\). By reducing the number of passages processed by the cross-encoder, we enable real-time computation while still taking advantage of its superior ability to capture inter-sentence relationships compared to a bi-encoder. However, passage retrievers that are pretrained on general text typically have difficulty handling financial tables. For instance, when retrieving tables, attributes such as titles, periods, and indicators are more important than the numerical values, yet these cues are not well captured. To address this, we fine-tuned the model on table data using FinQA training set where tables serve as evidence. Specifically, for each question \(q\) and its associated evidence document \(d\), we denote the tables on the evidence page as the evidence passage set \(\mathcal{X}\). For each evidence passage \(p \in \mathcal{X}\), we sample \(n_{neg}\) negative passages, where negative passages are defined as tables that appear on pages other than the evidence page. The cross-encoder is then trained with a binary cross-entropy loss~\cite{nogueira2019passage} defined as
\setlength{\abovedisplayskip}{5pt}
\setlength{\belowdisplayskip}{5pt}
\[
\footnotesize
\begin{aligned}
\mathcal{L} = \sum_{(q,p)\in \mathcal{X}} \Biggl[ & -\log\bigl(\text{CrossEncoder}^P(q,p)\bigr) \\
& - \sum_{p'\in \mathcal{P}^{-}} \log\Bigl(1-\text{CrossEncoder}^P(q,p')\Bigr) \Biggr],
\end{aligned}
\]
where the cross-encoder applies an internal sigmoid to produce scores in [0, 1].

\subsection{Evidence Curation}
\label{sec:evidencecuration}
Financial questions often involve comparisons between different time periods or companies. Even when the retrieval process selects relevant passages, some critical information may be missing.
Furthermore, retrieved passages can contain irrelevant data that hinders overall performance~\cite{liu2024lost, xu2024knowledge}. 
To overcome these issues, we introduce an evidence curation process that filters out incorrect data and fills information gaps by initiating additional retrieval when necessary. 
Our process comprises three modules: a passage filter that removes irrelevant passages (Section~\ref{sec:pf}), an answerability checker that assesses evidence sufficiency (Section~\ref{sec:ac}), and a complementary question generator that formulates a supplementary query if necessary (Section~\ref{sec:cqg}). We use an LLM to perform all three tasks in a single response (see Appendix~\ref{appendix:evidence_curation} for the prompt).

\subsubsection{Passage Filter} \label{sec:pf}
The passage filter removes passages from the retrieved set \(\mathcal{P}_{r}\) that are not relevant to the question, yielding a filtered passage set \(\mathcal{P}_{f}\) containing at most \(k_{P}'\) passages. The filter considers both newly retrieved passages and those previously identified as relevant from earlier iterations. This step is crucial, as the inclusion of noisy, irrelevant passages can lead the LLM to generate inaccurate responses.
\subsubsection{Answerability Checker} \label{sec:ac}
The answerability checker evaluates whether the filtered passages \(\mathcal{P}_{f}\) provide sufficient evidence to answer the question. If they are deemed adequate, \(\mathcal{P}_{f}\) and the question are forwarded to the answer generation stage; otherwise, the lack of sufficient information triggers a complementary iteration to retrieve additional data.

\subsubsection{Complementary Question Generator} \label{sec:cqg}
The complementary question generator examines the filtered passages \(\mathcal{P}_{f}\) to identify gaps in the evidence needed to answer the question. It then generates a supplementary query \(q_{c}\), which is used as input for the next retrieval.

\subsection{Answer Generation} \label{sec:ag}
In the answer generation stage, the relevant passages and the original question serve as inputs to a reasoning process that derives the final answer. 
For questions requiring numerical calculations, a Program-of-Thought (PoT) \cite{chen2022program} reasoning method is employed, while for text-based inferential questions, a Chain-of-Thought (CoT) \cite{wei2022chain} approach is applied. 
This dual strategy is particularly effective for financial documents, which are rich in numerical data and tables, ensuring comprehensive and accurate reasoning.
See Appendix~\ref{appendix:generation_prompt} for prompt details, adapted from DocMath-Eval~\cite{zhao-etal-2024-docmath}.

%% file: latex/5_experiments.tex
\input{tables/temp_tables}

\section{Experiments}
\label{experiments}
\vspace{-0.15cm}
\subsection{Experimental Settings}

\paragraph{Dataset.} 
We evaluate the open-domain QA methods on the \benchmark~benchmark proposed in Section~\ref{sec:benchmark}. 
All main experiments in this section are conducted on \benchmarkold, which contains 1,389 QA pairs.
Results on the updated benchmark \benchmarknew are reported separately in Appendix~\ref{appendix:expanded_lofin}.

For the purpose of the main experimental analysis, we categorize the question-answer pairs into three groups based on their format and context:
\begin{itemize}[noitemsep]
    \item \textbf{Numeric (Table):} The answer is a number from a table or can be calculated from numbers in tables.
    \item \textbf{Numeric (Text):} The answer is a number derived by extracting and combining numerical information from text.
    \item \textbf{Textual:} The answer is a textual explanation.
\end{itemize} 

Examples and statistics for these categories are in \autoref{tab:qa_distribution}.

\paragraph{Implementation details.}

Our system leverages several pre-trained and fine-tuned models across its components. 
For the passage retriever, we fine-tune a DeBERTa-v3 model~\cite{he2021debertav3} with \(n_{neg}=8\) for negative sampling, a batch size of 128, 3 epochs, and a learning rate of \(2 \times 10^{-7}\) on a single GeForce RTX 4090 GPU. For document retrieval, we employ the E5 model~\cite{wang2024multilingual} as the bi-encoder and use DeBERTa-v3 as the reranker. 
The answer generator is powered by OpenAI's GPT-4o, while other LLM-based tasks (query transformation, document summarization, and evidence curation) are handled by Qwen-2.5-7B-Instruct~\cite{yang2024qwen2}.

Our framework runs for a maximum of \(i_{\text{max}}=3\) iterations. The document retriever retrieves \(k_D=5\) documents, and the passage retriever extracts \(k_P=5\) passages, with the passage filter capping the output at \(k_{P}'=10\). Additional hyperparameters are provided in Appendix~\ref{appendix:experiment_hyperparams}.

\paragraph{Baseline methods.}
We compare our proposed method with several state-of-the-art RAG approaches, including RQ-RAG~\cite{chan2024rq} and Self-RAG~\cite{asai2023self}. 
To ensure a fair comparison with our approach—which employs GPT-4o as the answer generator—we use the same answer generator for the latest retrieval algorithms: Dense, HybridSearch~\cite{wang2024financerag}, HHR~\cite{arivazhagan-etal-2023-hybrid}, and IRCoT~\cite{trivedi2022interleaving}. 
In addition, we evaluate against Perplexity~\cite{Perplexity2023}, a commercial LLM service that combines web search with LLMs. 
RQ-RAG, Self-RAG, and IRCoT employ iterative LLM-based retrieval, while Dense serves as a strong baseline, it employs OpenAI's \texttt{text-embedding-3-small} as its encoder for DPR, with results reranked by DeBERTa-v3. 
All methods construct passages by concatenating the title and content during retrieval. Detailed configurations for each baseline are provided in Appendix~\ref{appendix:experiment_baseline}.

\paragraph{Metrics.}

We evaluate numeric answers for accuracy by considering rounding and truncation, while textual answers are evaluated using GPT-4o and FAMMA prompts \cite{xue2024famma} (see Appendix \ref{appendix:experiment_gpt_acc_prompt}). We measure retrieval performance at the page level, using recall and precision against ground-truth evidence pages as metrics. 
In particular, since chunk locations or units can vary for Page recall, we standardize them at the page level to ensure consistent performance measurement.

\subsection{Main Result}
\autoref{tab:main_results} shows the retrieval performance (page recall) and final answer accuracy for HiREC and the baselines.
Our approach outperforms all baselines, achieving at least 10\% higher page recall and 13\% higher answer accuracy than the second-best model, Dense. 
The result validates the effectiveness of our method in retrieval for standardized documents. 
 Furthermore, HiREC retrieves an average of only 3.7 passages, demonstrating its efficiency in selecting high-quality evidence through evidence curation.

Notably, in the textual category, the answer accuracies are higher than the page recalls for all methods. 
This suggests that LLMs can often answer text-based questions correctly even when retrieval is incomplete. 
Numeric questions require more precise reasoning and the table category remains especially challenging.

\begin{table}[!tb]
\centering
\footnotesize
\resizebox{\linewidth}{!}{
    \begin{tabular}{l|rr|r}
        \toprule
        \textbf{Method} & \makecell{\textbf{Page} \\ \textbf{Precision}} & \makecell{\textbf{Page} \\ \textbf{Recall}} & \makecell{\textbf{Ansswer} \\ \textbf{Accuracy}} \\
        \hline
        \textbf{HiREC}   & \textbf{21.79} & 45.35 & \textbf{42.36} \\
        ~~w/o HR  & 14.75 & 34.16 & 32.76 \\
        ~~w/o EC  & 4.70  & 41.41 & 36.70 \\
        ~~w/o Fine-tuning & 21.07 & 42.77 & 40.13 \\
        ~~w/o Filter   & 8.43     & \textbf{50.19} & 42.08     \\
        \bottomrule
    \end{tabular}

}
\caption{Ablation study}
\label{tab:ablation}
\end{table}

\subsection{Analysis}

\begin{figure}[!tb]
    \centering
    \setlength{\abovecaptionskip}{-5pt}
    \includegraphics[width=\linewidth]{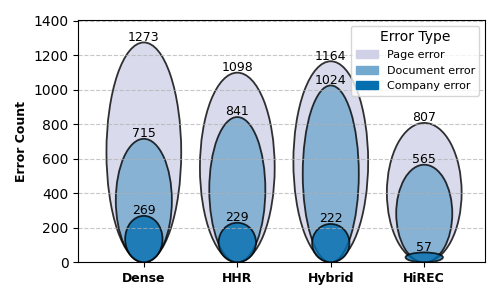}
    \caption{Comparison of company, document, page error rates for HiREC and baselines.}
    \label{fig:error_type_analysis}
\end{figure}

\paragraph{Ablation Study}
\autoref{tab:ablation} shows page precision, page recall and answer accuracy for HiREC when each component is removed.
Hierarchical retrieval (HR) and evidence curation (EC) denote the two components.
The setting w/o HR corresponds to the outcome of applying the Dense method in conjunction with EC, while the setting w/o EC represents the initial retrieval performance of HR (\(k_{P}=10\)).

The w/o Fine-tuning setting uses an unfine-tuned reranker to address fairness concerns related to table-specific prior knowledge. 
Despite a slight drop in performance, HiREC still outperforms the Dense baseline by over 10\% in accuracy. 
Even when Dense is paired with a fine-tuned reranker, it only reaches an average AnswerAcc of 30.55\%, maintaining a similar performance gap.

HR is critical for enhancing retrieval accuracy because its absence produces the lowest performance. 
In addition, the initial search performance of HR exceeds that of the dense method combined with EC.
The w/o EC setting demonstrates the effectiveness of both passage filter and the complementary question generator. 
HR supported by EC delivers enhanced precision, recall and answer accuracy. 
Furthermore, results from the w/o Filter setting reveal the advantages of the complement and filtering. 
Although the complementary component attains the highest recall score in the absence of filtering, its accuracy is lower than that of HiREC with filtering. 
This outcome is attributed to the inclusion of incorrect information that causes conflicts when filtering is not applied~\cite{xu2024knowledge}.

\begin{figure}[!tb]
    \centering
    \setlength{\abovecaptionskip}{-5pt}
    \includegraphics[width=\linewidth]{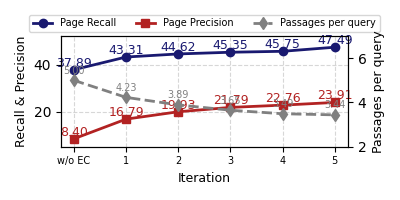}
    \caption{Recall, precision, and passages per query by iteration. EC stands for Evidence curation.}
    \label{fig:evidence_curation_by_iter}
\end{figure}

\paragraph{Error type analysis.} \autoref{fig:error_type_analysis} presents the error count for cases in which the company indicated in the document is incorrect for the retrieved passage of each method. A company error is defined as an instance in which both the passage and the document contain incorrect company information, and the results are reported based solely on the top-1 retrieved passage. HiREC achieves the fewest errors by correctly identifying companies during document retrieval, which in turn ensures accurate passage retrieval.

\begin{figure}[!tb]
    \centering
    \setlength{\abovecaptionskip}{-5pt}
    \includegraphics[width=0.8\linewidth]{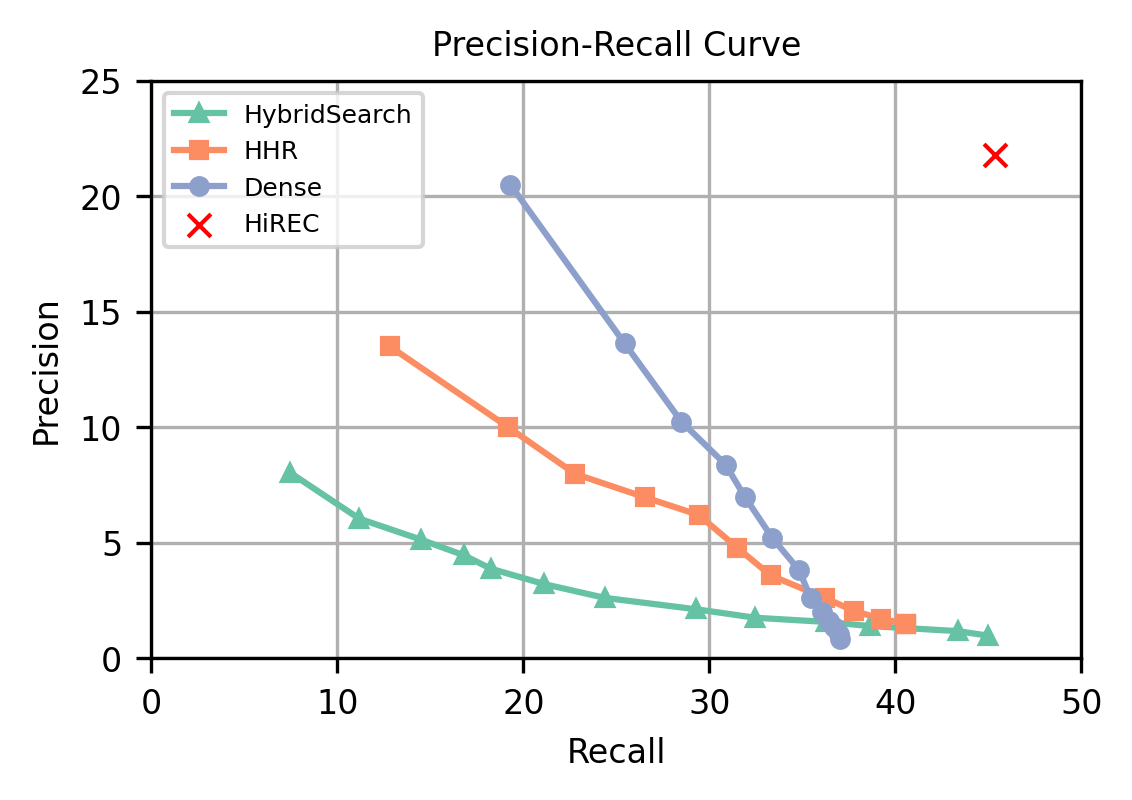}
    \caption{Precision-recall curve (recall on X-axis, precision on Y-axis)}
    \label{fig:pr_curve}
\end{figure}

\paragraph{Evidence curation by iteration.} \autoref{fig:evidence_curation_by_iter} shows that iterative evidence curation (EC) enhances page recall and precision while reducing passages per query compared to the initial hierarchical retrieval (w/o EC). 
As iterations progress, retrieval performance steadily improves and efficiency increases, demonstrating evidence curation effectiveness.

\paragraph{Precision-recall.} \autoref{fig:pr_curve} presents the precision-recall curve, where $k$ varies from 1 to 50 for baseline methods. As expected, increasing $k$ improves recall but lowers precision due to the retrieval of less relevant passages. \ours achieves both higher precision and recall, consistently exceeding the maximum values observed across all $k$ in the baseline range. This highlights ability of \ours to retrieve relevant information more effectively while maintaining accuracy.

\paragraph{Cost-efficiency.} 
\begin{table}[!tb]
\centering
\input{tables/efficiency_comparison}
    \label{table:efficiency_comp}
\end{table}
\autoref{table:efficiency_comp} presents the average input/output token counts and total API cost incurred during the retrieval and generation processes. 
HiREC achieves the highest performance while using significantly fewer tokens and lower cost than the baselines by filtering out irrelevant passages before the answer generation stage. 
Moreover, compared to IRCoT in retrieval, HiREC uses far fewer tokens, demonstrating that its filtering and complementary question generation enable efficient iteration. 
Finally, our results show that even a relatively small LLM can effectively perform curation without incurring high costs.

\input{latex/5-1_additional_experiments}
\input{tables/performance_results_by_data_source}

\begin{table*}[!]
  \centering
  \input{tables/case_study}
    \label{table:case_study}
\end{table*}
\input{latex/5-2_additional_experiment}
\begin{table}[!]
    \centering
    \input{tables/chatgpt}
\end{table}

\input{latex/5-3_case_study}

%% file: tables/temp_tables.tex
\begin{table*}[htbp]
\centering
\resizebox{\textwidth}{!}{
\begin{tabular}{l|rrrrrr|rrr}
\toprule
\multicolumn{1}{l|}{\textbf{Dataset}} & \multicolumn{2}{c}{\textbf{Numeric (Table)}} & \multicolumn{2}{c}{\textbf{Numeric (Text)}} & \multicolumn{2}{c|}{\textbf{Textual}} & \multicolumn{3}{c}{\textbf{Average}} \\
\cmidrule(lr){2-3} \cmidrule(lr){4-5} \cmidrule(lr){6-7} \cmidrule(lr){8-10}
\textbf{Method} & \makecell{\textbf{Page}\\\textbf{Recall}} & \makecell{\textbf{Answer}\\\textbf{Acc}} & \makecell{\textbf{Page}\\\textbf{Recall}} & \makecell{\textbf{Answer}\\\textbf{Acc}} & \makecell{\textbf{Page}\\\textbf{Recall}} & \makecell{\textbf{Answer}\\\textbf{Acc}} & \makecell{\textbf{Page}\\\textbf{Recall}} & \makecell{\textbf{Answer}\\\textbf{Acc}} & \makecell{{$\mathbf{k}$}} \\
\midrule
GPT-4o (Zero-shot)             & -      & 3.82   & -      & 2.93   & -      & 35.00  & -      & 13.92 &  - \\
Perplexity \cite{Perplexity2023}         & -      & 2.51   & -      & 5.13   & -      & 24.00  & -      & 10.55 & - \\
Self-RAG~\cite{asai2023self}    & 19.96  & 1.86   & 24.18  & 4.03   & 12.75   & 17.00  & 18.96  & 7.63 &  10.0 \\
RQ-RAG \cite{chan2024rq}  & 18.61  & 1.97   & 19.05  & 2.56   & 17.96  & 20.50  & 18.54  & 8.34 & 36.0 \\
IRCoT~\cite{trivedi2022interleaving}~$\lozenge$     & 28.17  & 19.10  & 34.62  & 27.84  & 12.67  & 20.00  & 25.15  & 22.31 & 20.0 \\
HybridSearch~\cite{wang2024financerag}~$\lozenge$ & 26.75  & 19.10  & 32.05  & 30.77  & 14.37  & 27.00  & 24.39  & 25.62 & 10.0 \\
HHR~\cite{arivazhagan-etal-2023-hybrid}~$\lozenge$       & 37.67  & 26.53  & 40.29  & 32.97  & 21.98  & 26.50  & 33.31  & 28.67 & 10.0 \\
Dense~\cite{karpukhin2020dense}~$\lozenge$       & 37.69  & 23.69  & 40.48  & 32.97  & 26.18  & 31.00  & 34.78  & 29.22 & 10.0 \\
HiREC (Ours)~$\lozenge$        & \textbf{50.17}  & \textbf{37.23}  & \textbf{53.48}  & \textbf{48.35}  & \textbf{32.39}  & \textbf{41.50}  & \textbf{45.35}  & \textbf{42.36} & 3.7\\
\midrule
Gold evidence~$\lozenge$ & 100    & 64.96  & 100   & 69.23  & 100   & 63.50  & 100    & 65.90 & 1.3 \\
\bottomrule
\end{tabular}
}
\caption{Main evaluation results on \benchmarkold. Methods marked with $\lozenge$ use GPT-4o for generation. 
Here, $k$ denotes the average number of input passages used during generation, and Gold evidence shows only the correct page. \textbf{Bold} indicates the highest performance.}
\label{tab:main_results}
\vspace{-0.5cm}
\end{table*}

%% file: tables/efficiency_comparison.tex
\centering
\small
\setlength{\tabcolsep}{10pt}
\renewcommand{\arraystretch}{1.2}
\begin{tabular}{@{}lrrrr@{}}
\toprule
\multirow{2}{*}{\textbf{Method}} & \multicolumn{2}{c}{\textbf{Retrieval}} & \multicolumn{2}{c}{\textbf{Generation}} \\ 
\cmidrule(lr){2-3} \cmidrule(lr){4-5}
                                & \textbf{Tokens} & \textbf{Cost (\$)}               & \textbf{Tokens} & \textbf{Cost (\$)}             \\ \midrule
Dense                           & -               & -                           & \makecell[r]{2,666 \\ / 177}  & \makecell[r]{9.3 \\ / 2.5}  \\
IRCoT                           & \makecell[r]{9,610 \\ / 1,372}  & \makecell[r]{33.4 \\ / 19.1} & \makecell[r]{3,475 \\ / 177}  & \makecell[r]{12.1 \\ / 2.5}  \\ \midrule
HiREC & \makecell[r]{4,291 \\ / 313} & \makecell[r]{$^*$14.9\\ / $^*$4.4} & \makecell[r]{1,052 \\ / 159} & \makecell[r]{3.7 \\ / 2.2} \\ \bottomrule
\end{tabular}
\caption{Cost-efficiency comparison. Each cell shows the values for the input/output. The asterisk (\(^*\)) indicates that, although a much smaller open-source LLM was used, the cost is computed using GPT-4o's pricing.}
\label{tab:cost_efficiency}

%% file: latex/5-1_additional_experiments.tex
\subsection{Analysis of Performance Across Various LLM Generators}

\begin{table}[!]
\centering
\resizebox{\linewidth}{!}{
\begin{tabular}{lrrrr}
\toprule
\textbf{Method} & \makecell{\textbf{Numeric} \\ \textbf{(Table)}} & \makecell{\textbf{Numeric} \\ \textbf{(Text)}} & \textbf{Textual} & \textbf{Average} \\
\midrule
\textbf{Dense} & \multicolumn{4}{c}{} \\
~~+ Qwen-2.5-7B & 19.76 & 22.34 & 29.52 & 23.87 \\
~~+ Deepseek-14B & 25.44 & 31.87 & 35.00 & 30.77 \\
~~+ GPT-4o & 23.69 & 26.18 & 31.00 & 29.22 \\
\hline
\textbf{\ours} & \multicolumn{4}{c}{} \\
~~+ Qwen-2.5-7B & \underline{27.62} & \underline{33.33} & \underline{36.00} & \underline{32.32} \\
~~+ Deepseek-14B & \underline{34.39} & \underline{41.39} & \underline{40.53} & \underline{38.76} \\
~~+ GPT-4o & \textbf{37.23} & \textbf{48.35} & \textbf{41.50} & \textbf{42.36} \\
\bottomrule
\end{tabular}
}
\caption{Performance comparison across various LLM generators. Underlined values denote HiREC configurations utilizing smaller generators that surpass Dense + GPT-4o performance.}
\label{tab:llm_generator_perf}
\end{table}

We thoroughly analyze the effectiveness of the \ours framework across diverse LLM generator models. For this analysis, additional evaluations were conducted using open-source models such as DeepSeek-R1-Distill-Qwen-14B\cite{guo2025deepseek} and Qwen-2.5-7B-Instruct\cite{yang2024qwen2} as generators, in place of GPT-4o.

As shown in \autoref{tab:llm_generator_perf}, \ours consistently demonstrates superior QA performance across various LLM generators, proving its robustness. 
Notably, even when employing smaller open-source models, \ours configurations outperform the Dense baseline. 
\ours +Deepseek-14B achieved over 9\% higher average answer accuracy compared to Dense+GPT-4o.
Another key implication is that integrating sLLMs in the retrieval stage proves to be an effective strategy for reducing overall inference costs while maintaining strong performance.

%% file: tables/performance_results_by_data_source.tex
\begin{table}[!]
\centering
\resizebox{\linewidth}{!}{
\begin{tabular}{lrrrrrr}
\toprule
\textbf{Method} & \multicolumn{2}{c}{\textbf{Financebench}} & \multicolumn{2}{c}{\textbf{FinQA}} & \multicolumn{2}{c}{\textbf{SEC-QA}} \\
\cmidrule(lr){2-3} \cmidrule(lr){4-5} \cmidrule(lr){6-7}
& \makecell{\textbf{Page} \\ \textbf{Recall}} & \makecell{\textbf{Answer} \\ \textbf{Acc}} & \makecell{\textbf{Page} \\ \textbf{Recall}} & \makecell{\textbf{Answer} \\ \textbf{Acc}} & \makecell{\textbf{Page} \\ \textbf{Recall}} & \makecell{\textbf{Answer} \\ \textbf{Acc}} \\
\midrule
Self-RAG & 9.00 & 13.33 & 22.93 & 3.24 & 4.59 & 4.72 \\
RQ-RAG & 13.67 & 18.67 & 20.32 & 2.88 & 9.38 & 4.72 \\
IRCoT & 9.33 & 20.00 & 32.24 & 22.66 & 4.20 & 7.09 \\
HybridSearch & 10.00 & 29.33 & 29.95 & 23.29 & 10.42 & 7.87 \\
HHR & 16.44 & 34.00 & 41.50 & 29.50 & 10.13 & 5.51 \\
Dense & 26.11 & 33.33 & 40.38 & 27.61 & 15.70 & 9.45 \\
\textbf{HiREC} & \textbf{40.00} & \textbf{50.00} & \textbf{52.49} & \textbf{41.61} & \textbf{19.97} & \textbf{14.17} \\
\bottomrule
\end{tabular}
}
\caption{Performance results by data source.}
\label{tab:performance_by_data_source}
\end{table}

%% file: tables/case_study.tex
\tiny
\renewcommand{\arraystretch}{0.6}
\setlength{\fboxsep}{0pt}

\resizebox{\textwidth}{!}{%
\begin{tabular}{m{1.2cm} p{10cm}}
\toprule
\raggedright \textbf{Step} & \textbf{Content} \\
\midrule
\raggedright \textbf{Question} & 
\textit{What is Adobe's year-over-year change in unadjusted operating income from \textbf{\textcolor{orange}{FY2015}} to \textbf{\textcolor{purple}{FY2016}} (in units of percents and round to one decimal place)? Give a solution to the question by using the income statement.} \\
\midrule
\raggedright \textbf{Initial Retrieval} & 
\textbf{\textcolor{green}{\cmark} ADBE\_2015\_10K (p.59): }3,112,300 3,045,960 \fcolorbox{blue!50}{blue!10}{\strut Operating income 903,095} 412,685 422,723 Non-operating  ...  \\

& \textbf{\textcolor{red}{\xmark} \;ADBE\_2015\_10K (p.39): }Net income of \$629.6 million increased by \$361.2 million, or 135\%, during fiscal 2015 ... \vspace{3pt} \\

& \textbf{\textcolor{red}{\xmark} \;ADBE\_2015\_10K (p.45):} Amortization of purchased intangibles 68.7 ... Total operating expenses \$3,148.1 ... \vspace{3pt} \\
 


\midrule

\raggedright \textbf{Evidence \\Curation} &

\textbf{Relevant passages:} ADBE\_2015\_10K (p.59) \vspace{3pt} \quad \textbf{Answerable:} False \vspace{3pt}\\

& \textbf{Explanation:} The context does not provide operating income figures for both \textbf{\textcolor{orange}{FY2015}} and \textbf{\textcolor{purple}{FY2016}}. We only have operating income for \textbf{\textcolor{orange}{FY2015}} and some intermediate values, but no direct comparison between \textbf{\textcolor{orange}{FY2015}} and \textbf{\textcolor{purple}{FY2016}}. \vspace{3pt} \\

& \textbf{Complementary question:} \textit{What is Adobe's operating income for \textbf{\textcolor{purple}{FY2016}}?} \\
\midrule
\raggedright \textbf{Complementary Retrieval} & 
\textbf{\textcolor{green}{\cmark} ADBE\_2016\_10K (p.61): }3,148,099 3,112,300 \fcolorbox{blue!50}{blue!10}{\strut Operating income 1,493,602} 903,095 412,685 Non-operating ... \vspace{3pt} \\

& \textbf{\textcolor{red}{\xmark} \;ADBE\_2016\_10K (p.35):} ITEM 6. SELECTED FINANCIAL DATA  ... Net income \$1,168,782 \$ 629,551 ... \vspace{3pt} \\

& \textbf{\textcolor{red}{\xmark} \;ADBE\_2017\_10K (p.57):} CONSOLIDATED STATEMENTS OF INCOME ... Operating income 2,168,095 ... \vspace{3pt} \\



\midrule
\raggedright \textbf{Evidence \\Curation} &

\textbf{Relevant passages:}
ADBE\_2015\_10K (p.59), ADBE\_2016\_10K (p.61) \quad \textbf{Answerable:} True \vspace{3pt} \\
\midrule
\raggedright \textbf{Generation} & 
\textbf{Ground Truth Answer}: 46\% \quad \textbf{Generated Answer}: 46 \quad \textbf{Correct:} \textcolor{green}{\cmark} \\
\bottomrule
\end{tabular}
}
\caption{Case study illustration of HiREC framework effectiveness}
\vspace{-0.5cm}
\label{tab:two-col-table}

%% file: latex/5-2_additional_experiment.tex

\subsection{Performance results by data source}
Our analysis of various data sources aims to evaluate benchmark data leakage risk and demonstrate the robustness of the \ours framework.

As shown in \autoref{tab:performance_by_data_source}, \ours consistently demonstrates superior performance across all data sources. 
Notably, all methods, including \ours, exhibit relatively lower performance on the SEC-QA subset, as multi-document QA tasks require the synthesis of information from multiple sources for accurate answers.
These results suggest the \benchmark~benchmark is more challenging due to multi-document scenarios and mitigates leakage risk by relying on retrieved evidence.

The final results for the \benchmark~benchmark, including the expanded SEC-QA subset, can be found in the Appendix.

\subsection{Comparison of Commercial LLMs with Web Search}

Commercial systems such as SearchGPT and Perplexity combine LLMs with web search to answer questions using financial data. \autoref{tab:searchgpt_results} compares these systems with our approach where Perplexity uses the llama-3.1-sonar-large-128k-online model and SearchGPT uses GPT-4o. In our evaluation based on 40 questions per category, HiREC consistently outperforms these baselines especially on numeric questions, indicating that although commercial systems effectively retrieve relevant documents they often miss the precise numerical details necessary for accurate computation.

%% file: tables/chatgpt.tex
\small
\setlength{\tabcolsep}{4pt}
\begin{tabular}{lrrrr}
\toprule
    \makecell{\textbf{Method}} & \makecell{\shortstack[c]{\textbf{Numeric}\\\textbf{(Table)}}} & \makecell{\shortstack[c]{\textbf{Numeric}\\\textbf{(Text)}}} & \makecell{\textbf{Textual}} & \makecell{\textbf{Average}} \\
    \midrule
    Perplexity & 2.5  & 5.0  & 37.5 & 15.0 \\
    SearchGPT  & 15.0 & 32.5 & 35.0 & 27.5 \\
    \midrule
    HiREC     & 30.0 & 65.0 & 45.0 & 46.7 \\
\bottomrule
\end{tabular}
\caption{Answer accuracy for Perplexity, SearchGPT, and HiREC on 40 samples per category.}

\label{tab:searchgpt_results}

%% file: latex/5-3_case_study.tex
\subsection{Case Study}
We analyze how evidence curation filters out unnecessary information and effectively retrieves missing details, and how this impacts the final results. \autoref{table:case_study} presents the retrieval and evidence curation outcomes over iterations for a given question. The initial pass of hierarchical retrieval retrieves passages containing operating income for Adobe for FY2015, but the question remains unanswerable due to missing FY2016 data. The answerability checker detects this gap and triggers the complementary question generator, leading to a complementary pass that retrieves the missing FY2016 operating income. With complete evidence, the answer generator successfully computes the 46\% year-over-year change, highlighting the effectiveness of iterative refinement in HiREC.

%% file: latex/6_conclusion_and_others.tex
\section{Conclusion}

We introduced HiREC, a retrieval-augmented framework for question answering over standardized financial documents. Our hierarchical retrieval reduces confusion from repeated boilerplate content, while evidence curation filters irrelevant passages and recovers missing information.  
To evaluate under realistic open-domain conditions, we constructed \benchmark, a large-scale benchmark with 1,595 QA pairs across 145,000 SEC filings. The dataset includes multi-document and multi-hop questions that go beyond prior financial QA benchmarks.  
Experiments show that HiREC consistently outperforms state-of-the-art baselines in both retrieval quality and answer accuracy, while maintaining cost efficiency through selective passage use.  
Overall, our findings suggest that HiREC provides a scalable and effective solution for open-domain QA over complex, standardized financial documents.

%% file: latex/8_limitation.tex
\section*{Limitations}
In our study, we use LLMs for query transformation, evidence curation, and answer generation, making our approach dependent on the performance of the LLMs. We utilize a relatively small LLM, Qwen 2.5 7B~\cite{yang2024qwen2}, compared to commercial models.

The benchmarks we employ, FinQA and FinanceBench, are publicly available datasets. Therefore, it is possible that pre-trained large language models have already been exposed to these datasets during their training. In \autoref{tab:main_results}, GPT-4o (Zero-shot) achieves the second-highest textual performance.

\section*{Acknowledgements}
This study was supported by the National R\&D Program for Cancer Control through the National Cancer Center (NCC), funded by the Ministry of Health \& Welfare, Republic of Korea (RS-2025-02264000).

This work was also supported by the Institute of Information \& Communications Technology Planning \& Evaluation (IITP) grant funded by the Korea government (MSIT) (No. RS-2023-00261068, Development of Lightweight Multimodal Anti-Phishing Models and Split-Learning Techniques for Privacy-Preserving Anti-Phishing), and by the Convergence Security Core Talent Training Program (IITP-2024-RS-2024-00423071), supervised by IITP and also funded by the Ministry of Science and ICT (MSIT), Korea.

%% file: latex/7_appendix.tex
\newpage
\section*{Appendix}
\appendix 

\input{latex/appendix/experiments_expanded_lofin}

\section{Supplementary Details on the \benchmark~Benchmark}
\subsection{Categorization Details} \label{appendix:experiment_data}
Tables~\ref{tab:qa_distribution} and~\ref{tab:expaned_qa_distribution} summarize the distribution of QA pairs in \benchmark~by answer type before and after the SEC-QA expansion. The original dataset is heavily skewed toward numeric-table questions, while the expanded version introduces a larger portion of textual and multi-hop reasoning questions, illustrating the increased complexity and diversity of the benchmark.

\begin{table}[ht]
    \centering
    \small
    \begin{tabular}{lcccc}
        \toprule
        \textbf{Category} & \makecell{\textbf{Numeric} \\ \textbf{(Text)}} & \makecell{\textbf{Numeric} \\ \textbf{(Table)}} & \textbf{Textual} & \textbf{Total} \\
        \midrule
        \textbf{\# QAs} & 273 & 916 & 200 & 1389 \\
        \textbf{Ratio (\%)} & 19.65 & 65.95 & 14.40 & 100 \\
        \bottomrule
    \end{tabular}
    \caption{Distribution of initial \benchmark~QA pairs by category.}
    \label{tab:qa_distribution}
\end{table}

\begin{table}[ht]
    \centering
    \small
    \begin{tabular}{lcccc}
        \toprule
        \textbf{Category} & \makecell{\textbf{Numeric} \\ \textbf{(Text)}} & \makecell{\textbf{Numeric} \\ \textbf{(Table)}} & \textbf{Textual} & \textbf{Total} \\
        \midrule
        \textbf{\# QAs} & 273 & 925 & 397 & 1595 \\
        \textbf{Ratio (\%)} & 17.11 & 58.00 & 24.89 & 100 \\
        \bottomrule
    \end{tabular}
    \caption{Distribution of expanded \benchmark~QA pairs by category.}
    \label{tab:expaned_qa_distribution}
\end{table}

\subsection{FinQA Open-Domain Conversion Prompt} \label{appendix:bench_finqa_prompt}

\begin{table}[H]
    \small
    \centering
    \renewcommand{\arraystretch}{1.2}
    \resizebox{\columnwidth}{!}{
    \begin{tabular}{|p{0.9\linewidth}|}
    \hline
    You are a financial AI Assistant. The following financial questions are provided without including the company names and document period. Rewrite the questions to include the given information.\\[0.5ex]
    \\
    - Maintain the original meaning of the question while allowing for varied expressions that enable accurate and open-ended responses without altering the factual content.\\[0.5ex]
    - The question must include the company name and the report year.\\[0.5ex]
    - Ticker is used as information about the company name and should not be included in the question.\\[0.5ex]
    - While rephrasing the question, integrate additional information naturally into the sentence without using simple connectors like 'In', 'for', or 'according to'.\\[0.5ex]
    \\
    Document information:\\[0.5ex]
    \quad - company ticker: \{ticker\}\\[0.5ex]
    \quad - document\_period: \{period\}\\[0.5ex]
    \\
    Question: \{question\}\\[0.5ex]
    \\
    Output format is \#\#new\_question:\\[0.5ex]
    \hline
    \end{tabular}
    }
    \caption{This prompt template is used to convert closed-domain FinQA questions into an open-domain format by seamlessly integrating company names and report years into the questions, while preserving their original meaning.}
    \label{tab:bench_finqa_prompt}
\end{table}

\subsection{FinQA Gold Page Selection} \label{appendix:bench_finqa_anno}

Our evidence page annotation for FinQA relies on a dual-method approach. Using FinQA's qid, we first identify candidate documents and discard any questions associated with documents that have page separation issues. Within each candidate document, we compute BM25 similarity between the FinQA gold table context (which emphasizes distinct numerical values) and each page's content to select candidate evidence pages. Additionally, an NLI model evaluates the semantic similarity between the concatenated pre\_text and post\_text (i.e., the context surrounding the table) and the content of each page. If both methods select the same top candidate, that page is accepted and subsequently verified; otherwise, manual annotation is performed to ensure accuracy. Out of 1147 instances, 894 (78\%) were automatically accepted. However, among these accepted pages, 9 instances (approximately 1\% of 894) contained errors due to OCR issues that merged content from two pages into one. In the remaining 218 instances (19\%), the top pages selected by the two methods differed, necessitating manual annotation, and an additional 35 cases (3\%) were discarded during preprocessing. This combined approach ensures robust and accurate identification of the correct evidence pages for FinQA.

\subsection{Question Templates of SEC-QA} \label{appendix:bench_secqa}

\begin{itemize} [noitemsep]
    \item How much common dividends did \{company\} pay in the last \{num\_year\} years in US dollars?
    \item What is the percentage difference of \{company1\}’s \{metric\} compared to that of \{company2\}?
    \item What is \{company\}’s overall revenue growth over the last \{num\_year\}-year period?
    \item Among \{company\_names\}, what is the \{metric2\} of the company that has the highest \{metric1\}?
\end{itemize}

\section{Experimental Settings}

\subsection{Hyperparameters}
\label{appendix:experiment_hyperparams}
\begin{itemize} [noitemsep]
    \item Candidate document count \(k_{D}': 100\)
    \item Final document count \(k_{D}: 5\)
    \item Final passage count \(k_{P}: 5\)
    \item Maximum relevant passages \(k_{P}'\): 10
    \item Maximum iterations \(i_{max}: 3\)
    \item LLM temperature: 0.01
    \item Chunking tool: Langchain\footnote{\url{https://www.langchain.com/}}'s RecursiveCharacterTextSplitter
    \item Chunk size: 1024
    \item Overlap between chunks: 30
\end{itemize}

\subsection{Baselines}
\label{appendix:experiment_baseline}

\textbf{Perplexity:} The Perplexity baseline employs the \texttt{llama-3.1-sonar-large-128k-online} model, integrating web search to supplement the context used for answer generation.

\textbf{Self-RAG:} Self-RAG utilizes a 13B model in a Long-form setting. It leverages the Contriever model to retrieve 10 context passages for each query, ensuring comprehensive coverage of the relevant information.

\textbf{RQ-RAG:} \texttt{text-embedding-3-small} is used in RQ-RAG to retrieve three context passages at each retrieval step. Configured for multi-hop QA with an exploration depth of two, it employs iterative retrieval to continuously refine the contextual information used for answer generation.

\textbf{HybridSearch:}
HybridSearch combines BM25-based scores with dense scores computed using the \texttt{text-embedding-3-small} model. This hybrid approach adheres to its original configuration to effectively balance lexical and semantic matching.

\textbf{IRCoT:}
IRCoT relies on a BM25-based retriever and employs GPT-4o for chain-of-thought (CoT) reasoning. It performs iterative retrieval by executing up to 3 iterations, with 5 passages being retrieved during each iteration.

\textbf{HHR:}
HHR implements a hybrid retrieval strategy that combines dense and sparse retrieval methods. In this approach, the \texttt{text-embedding-3-small} model is used for dense retrieval, and the system first retrieves the top 5 documents; within each document, the top 10 passages are then selected.

\textbf{Dense:}
The Dense method uses the \texttt{text-embedding-3-small} model to initially retrieve the top 50 relevant passages. These passages are subsequently reranked using a DeBERTa-v3 model to determine their final order.

\subsection{Textual Evaluation Prompt} \label{appendix:experiment_gpt_acc_prompt}
This prompt is used to evaluate the factual correctness of generated textual answers by instructing the LLM to compare them against the ground truth within the provided context.

\begin{table}[H]
    \small
    \centering
    \label{tab:gpt_acc_prompt_template}
    \begin{tabular}{|p{0.9\linewidth}|}
    \hline
    \vspace{0.5em} \\ 
    You are a highly knowledgeable expert and teacher in the finance domain.\\[0.5ex]
    You are reviewing a student’s answers to financial questions.\\[0.5ex]
    You are given the context, the question, the student’s answer and the student’s explanation and the ground\,-\,truth answer.\\[0.5ex]
    Please use the given information and refer to the ground\,-\,truth answer to determine if the student’s answer is correct.\\[0.5ex]
    The input information is as follows:\\[0.5ex]
    \quad context: '\{gold context\}'\\[0.5ex]
    \quad question: '\{question\}'\\[0.5ex]
    \quad ground-truth answer: '\{answer\}'\\[0.5ex]
    \quad student’s answer: '\{generated answer\}'\\[0.5ex]
    Please respond directly as either 'correct' or 'incorrect'. \\
    \vspace{0.5em} \\ 
    \hline
    \end{tabular}
    \caption{GPT-4o LM Evaluation Prompt Template}
\end{table}





\input{latex/appendix/prompt_list}
\input{latex/appendix/legalbenchmark}
\input{latex/appendix/case_study_failure}

\newpage

\begin{table*}[ht]
    \small
    \centering
    \input{tables/category}
\end{table*}

%% file: latex/appendix/experiments_expanded_lofin.tex
\section{Experiments on the Expanded \benchmarknew Benchmark}
\label{appendix:expanded_lofin}

\subsection{Motivation and Dataset Extension}
\label{appendix:dataset_distribution}
The initial version of the \benchmarkold benchmark was heavily biased toward single-document QA, particularly from FinQA, which limited its coverage of complex financial analysis scenarios. 
To address this limitation and incorporate reviewer feedback, we expanded the SEC-QA subset from 127 to 333 question-answer pairs by adding 206 new examples.
All newly added QA pairs were constructed from recent SEC filings collected up to April 2025, and are designed to require multi-document and multi-page reasoning. 
The questions include inter-company comparisons, trend analysis, and temporal reasoning. 
All examples were manually created, which prevents the possibility of data leakage.

\begin{table}[ht]
    \centering
    \resizebox{\linewidth}{!}{
    \begin{tabular}{lrrrr}
        \toprule
        \textbf{Dataset Source} & \textbf{FinQA} & \textbf{SEC-QA} & \textbf{Financebench} & \textbf{Total} \\
        \midrule
        \textbf{\# QAs} & 1,112 & 333 & 150 & 1,595 \\
        \textbf{Ratio (\%)} & 69.72 & 20.88 & 9.40 & 100 \\
        \bottomrule
    \end{tabular}
    }
    \caption{Distribution of expanded \benchmarknew QA pairs by dataset source.}
    \label{tab:qa_source_distribution}
\end{table}

\subsubsection{Experiments}
\label{appendix:experiments}
This section reports the results on the expanded benchmark \benchmarknew, following the same settings as in Section~\ref{experiments}.
The evaluation compares our proposed model \ours against the second-best baseline, Dense.
\autoref{tab:expanded_main_results} presents the main results by answer type. 
\ours consistently outperforms Dense across all categories.  
\autoref{tab:expanded_source_results} shows performance by data source. 
Although SEC-QA is a more challenging task, \ours still demonstrates superior results compared to the baseline.
The category distribution by answer type is summarized in \autoref{tab:expaned_qa_distribution}, and the dataset composition by source is reported in \autoref{tab:qa_source_distribution}.

\subsubsection{Discussion and Conclusion}
The experimental results on the extended SEC-QA subset in expanded \benchmarknew validate the robustness and effectiveness of \ours under realistic multi-document and multi-page QA settings.
With this expansion, the \benchmarknew benchmark now represents a significantly more challenging and realistic task compared to existing financial QA datasets.
Overall, these results demonstrate that \ours effectively performs evidence-based reasoning and provide strong empirical support that addresses reviewer concerns regarding dataset diversity, integrity, and model generalization.

\begin{table*}[!]
\centering
\resizebox{\textwidth}{!}{
\begin{tabular}{l *{3}{c} *{3}{c} *{3}{c} *{3}{c}}
\toprule
\textbf{Model} & \multicolumn{3}{c}{\textbf{Numeric Table}} & \multicolumn{3}{c}{\textbf{Numeric Text}} & \multicolumn{3}{c}{\textbf{Textual}} & \multicolumn{3}{c}{\textbf{Average}} \\
\cmidrule(lr){2-4} \cmidrule(lr){5-7} \cmidrule(lr){8-10} \cmidrule(lr){11-13}
& Precision & Recall & Accuracy & Precision & Recall & Accuracy & Precision & Recall & Accuracy & Precision & Recall & Accuracy \\
\midrule
Dense & 3.47 & 33.88 & 22.92 & 3.77 & 37.73 & 31.50 & 2.90 & 16.18 & 23.93 & 3.38 & 29.27 & 26.12 \\
\ours & \textbf{25.63} & \textbf{48.76} & \textbf{35.57} & \textbf{21.64} & \textbf{52.75} & \textbf{44.32} & \textbf{13.54} & \textbf{23.59} & \textbf{26.95} & \textbf{20.27} & \textbf{41.70} & \textbf{35.61} \\
\bottomrule
\end{tabular}
}
\caption{Performance by answer type on the expanded \benchmarknew benchmark.}
\label{tab:expanded_main_results}
\end{table*}

\begin{table*}[!]
\centering
\resizebox{\textwidth}{!}{
\begin{tabular}{l *{3}{c} *{3}{c} *{3}{c} *{3}{c}}
\toprule
\textbf{Model} & \multicolumn{3}{c}{\textbf{FinanceBench}} & \multicolumn{3}{c}{\textbf{FinQA}} & \multicolumn{3}{c}{\textbf{SEC-QA}} & \multicolumn{3}{c}{\textbf{Average}} \\
\cmidrule(lr){2-4} \cmidrule(lr){5-7} \cmidrule(lr){8-10} \cmidrule(lr){11-13}
& Precision & Recall & Accuracy & Precision & Recall & Accuracy & Precision & Recall & Accuracy & Precision & Recall & Accuracy \\
\midrule
Dense & 1.80 & 16.00 & 36.00 & 4.00 & 37.28 & 26.44 & 2.88 & 12.66 & 13.51 & 2.81 & 21.98 & 25.00 \\
\ours & \textbf{15.67} & \textbf{37.33} & \textbf{47.33} & \textbf{25.44} & \textbf{51.18} & \textbf{39.44} & \textbf{12.74} & \textbf{18.67} & \textbf{14.85} & \textbf{17.95} & \textbf{35.73} & \textbf{33.63} \\
\bottomrule
\end{tabular}
}
\caption{Performance by data source on the expanded \benchmarknew benchmark.}
\label{tab:expanded_source_results}
\end{table*}

%% file: latex/appendix/prompt_list.tex
\section{LLM Prompts} \label{appendix:prompt_list}
This section presents a detailed overview of the prompting strategies employed in our framework. Our prompt templates are designed based on common instruction formats introduced in the Prompt Engineering Guide\footnote{\url{https://www.promptingguide.ai/}}, and were initially generated using GPT-4o in an instruction-following mode. To ensure consistency with each module’s output format, we manually refined the initial prompts through a lightweight post-editing process.

No explicit constraints were imposed during prompt construction. We did not apply few-shot prompting due to practical limitations such as prompt length and inference cost. The final prompts are tailored to the specific needs of each module—query rewriting, document summarization, evidence curation, and answer generation—and are provided in this section.

These prompt designs play a central role in facilitating efficient information processing and enabling accurate responses to complex financial questions. By disclosing our prompting approach in detail, we aim to enhance the reproducibility and transparency of our overall system.

\subsection{Summarization Prompt} \label{appendix:summarization_prompt}
To assist document indexing during hierarchical retrieval (section \ref{document_retriever}), we use a concise LLM-based summarization prompt as follows \autoref{prompt:summary}:

\begin{table}[H]
    \small
    \centering
    \begin{tabular}{|p{0.9\linewidth}|}
    \hline
    \vspace{0.5em} \\ 
    You are a helpful assistant.
    Summarize the following text:
    \\
    \vspace{0.5em} \\ 
    \hline
    \end{tabular}
    \caption{Summarization prompt}
    \label{prompt:summary}
\end{table}
\newpage

\subsection{Query Transformation Prompt}
\label{appendix:query_transform_prompt}
To reduce noise from overly specific or ambiguous financial terms in user queries, we rewrite them into focused retrieval queries using the prompt below \autoref{prompt:transform_prompt}:

\begin{table}[H]
    \small
    \centering
    \begin{tabular}{|p{0.9\linewidth}|}
    \hline
    \vspace{0.5em} \\ 
    You are an AI that rewrites user questions about financial topics into concise meta-focused queries. \\
    \\
    1) Identify the key financial terms or metrics in the question. \\
    2) Determine which type of documents typically contain those terms. \\
    3) Transform the user’s question into a short query referencing the financial terms and the relevant documents. \\
    4) Do not reveal the transformation process or provide examples. \\
    5) Output only the final rewritten query. \\
    \\
    \#\# Question: \{Question\} \\
    \\
    \#\#\# Output format \\
    \#\# Query: \{Rewritten query\} \\
    \vspace{0.5em} \\ 
    \hline
    \end{tabular}
    \caption{Financial query transformation instructions}
    \label{prompt:transform_prompt}
\end{table}

\subsection{Evidence Curation Prompt} \label{appendix:evidence_curation}
The evidence curation module jointly performs three tasks—passage filtering, answerability checking, and complementary question generation—using a single prompt as follows \autoref{prompt:evidence_curation}:

\begin{table}[H]
    \small
    \centering
    \begin{tabular}{|p{0.9\linewidth}|}
    \hline
    \vspace{0.5em} \\ 
    \#\#\# Instruction \\
    You are a financial expert. Evaluate the provided context to determine if it contains enough information to answer the given question. \\
    1. Read the context carefully and decide if it contains enough information to answer the question. \\
    2. If it is answerable, set 'is\_answerable: answerable' and provide the answer in 'answer'. \\
    3. If it is not answerable, set 'is\_answerable: unanswerable'. Then: \\
        - List the relevant document IDs in 'answerable\_doc\_ids' in order of relevance (from most to least relevant). \\
        - Explain what specific information is missing in 'missing\_information'. \\
        - Provide a concise question in 'refined\_query' to search for exactly that missing information. \\
    4. Output your result strictly in the specified format below using '\#\#' headers. \\
    \\
    \#\#\#Inputs \\
    Context: \\
    Context1 (ID: 1): Title is \{title1\}. Content is \{content1\} \\
    Context2 (ID: 2): Title is \{title2\}. Content is \{content2\} \\
    ... \\
    Question: \{question\} \\
    \\
    \#\#\# Output format \\
    \#\# is\_answerable: answerable or unanswerable \\
    \#\# missing\_information: If 'unanswerable', specify the details or data needed; if 'answerable', None \\
    \#\# answer: If 'answerable', provide the answer; if 'unanswerable', then None \\
    \#\# answerable\_doc\_ids: Provide a list of document IDs that contain relevant information (e.g., [1, 2]). If none, use [] \\
    \#\# refined\_query: If 'unanswerable', provide a refined question to obtain the missing information; \\
    \vspace{0.5em} \\ 
    \hline
    \end{tabular}
    \caption{Evidence curation instructions}
    \label{prompt:evidence_curation}
\end{table}

\subsection{Generation Prompt} \label{appendix:generation_prompt}
We adopt two generation styles depending on the question type. 
For numerical reasoning tasks, we use Program-of-Thoughts (PoT) prompting to generate Python code (see \autoref{prompt:python_program}); 
for textual reasoning, we apply Chain-of-Thought (CoT) prompting (see \autoref{prompt:cot_prompt}).

\begin{table}[H]
    \small
    \centering
    \begin{tabular}{|p{0.9\linewidth}|}
    \hline
    \vspace{0.5em} \\ 
    \textbf{[System Input]} \\
    You are a financial expert, you are supposed to generate a Python program to answer the given question based on the provided financial document context. The returned value of the program is supposed to be the answer. \\
    \vspace{0.5em} \\
    \texttt{\`{}\`{}\`{}python} \\
    \texttt{def solution():} \\
    \texttt{\quad \# Define variables name and value based on the given context} \\
    \texttt{\quad guarantees = 210} \\
    \texttt{\quad total\_exposure = 716} \\
    \texttt{} \\
    \texttt{\quad \# Do math calculation to get the answer} \\
    \texttt{\quad answer = (guarantees / total\_exposure) * 100} \\
    \texttt{} \\
    \texttt{\quad \# return answer} \\
    \texttt{\quad return answer} \\
    \texttt{\`{}\`{}\`{}} \\
    \vspace{0.5em} \\
    \textbf{[User Input]} \\
    Context: \\
    Sources: \{title1\} - \{content1\} \\
    Sources: \{title2\} - \{content2\} \\
    ...\\
    \\
    Question: \{question\} \\
    \\
    Please generate a Python program to answer the given question. The format of the program should be the following: \\
    \vspace{0.5em} \\
    \texttt{\`{}\`{}\`{}python} \\
    \texttt{def solution():} \\
    \texttt{\quad \# Define variables name and value based on the given context} \\
    \texttt{\quad ...} \\
    \texttt{\quad \# Do math calculation to get the answer} \\
    \texttt{\quad ...} \\
    \texttt{\quad \# return answer} \\
    \texttt{\quad return answer} \\
    \texttt{\`{}\`{}\`{}} \\
    \vspace{0.5em} \\
    Continue the program to answer the question. The returned value of the program is supposed to be the answer: \\
    \vspace{0.5em} \\
    \texttt{\`{}\`{}\`{}python} \\
    \texttt{def solution():} \\
    \texttt{\quad \# Define variables name and value based on the given context} \\
    \texttt{\`{}\`{}\`{}} \\
    \vspace{0.5em} \\ 
    \hline
    \end{tabular}
    \caption{The prompt for generating a Python program to answer a financial question.}
    \label{prompt:python_program}
\end{table}

\begin{table}[H]
    \small
    \centering
    \begin{tabular}{|p{0.9\linewidth}|}
    \hline
    \vspace{0.5em} \\ 
    \textbf{[System Input]} \\
     You are a financial expert, and you are supposed to answer the given question based on the provided financial document context. You need to first think through the problem step by step, documenting each necessary step. Then, you are required to conclude your response with the final answer in your last sentence as "Therefore, the answer is {final answer}. 
    \vspace{0.5em} \\
    \textbf{[User Input]} \\
    Context: \\
    Sources: \{title1\} - \{content1\} \\
    Sources: \{title2\} - \{content2\} \\
    ...\\
    \\
    Question: \{question\} \\
    \\
    Let's think step by step to answer the given question.\\
    \vspace{0.5em} \\ 
    \hline
    \end{tabular}
    \caption{Chain-of-Thought prompt for generating a Python program to answer a financial question.}
\label{prompt:cot_prompt}
\end{table}

%% file: latex/appendix/legalbenchmark.tex
\section{Generalization to Other Domains}
\input{tables/legalbench}


To evaluate the effectiveness of our method beyond the financial domain, we assess retrieval performance on the LegalBench-RAG \cite{pipitone2024legalbench} dataset, a legal-domain benchmark where retrieval is particularly challenging due to domain-specific terminology and structured documents. Using a mini question set, we conduct retrieval over the full document corpus to simulate realistic conditions.

As shown in \autoref{table:legalbench}, \ours achieves the highest Precision@k (41.42) and Recall@k (48.50) while using the lowest average number of retrieved passages per query ($k=3.2$). This result highlights the framework’s ability to efficiently retrieve essential content with minimal redundancy. In particular, \ours demonstrates strong performance on structured datasets such as MAUD and CUAD, outperforming other models by a large margin in both P@k and R@k.

ContractNLI is a document-level natural language inference task in which each contract is evaluated against a fixed set of hypotheses. Unlike passage-based QA tasks, it requires global reasoning over the full document. As a result, Dense may benefit from broad recall, potentially retrieving passages that touch on relevant hypotheses by chance. In contrast, \ours follows a structured evidence curation pipeline that emphasizes precision, often selecting fewer but more targeted documents. This structural difference helps explain the smaller performance gap between \ours and Dense on ContractNLI compared to other tasks.

Overall, these results confirm the generalizability of \ours to non-financial domains, demonstrating robustness across diverse legal document types and retrieval tasks.

%% file: tables/legalbench.tex

\begin{table*}[!]
\centering
\small
\resizebox{\textwidth}{!}{
\begin{tabular}{lccccccccccc}
\toprule
\multirow{2}{*}{\textbf{Model}} &
\multicolumn{2}{c}{\textbf{PrivacyQA}} &
\multicolumn{2}{c}{\textbf{ContractNLI}} &
\multicolumn{2}{c}{\textbf{MAUD}} &
\multicolumn{2}{c}{\textbf{CUAD}} &
\multicolumn{2}{c}{\textbf{Average}} &
\multirow{2}{*}{\textbf{Avg. $k$}} \\
\cmidrule(lr){2-3} \cmidrule(lr){4-5} \cmidrule(lr){6-7} \cmidrule(lr){8-9} \cmidrule(lr){10-11}
& P@k & R@k & P@k & R@k & P@k & R@k & P@k & R@k & P@k & R@k \\
\midrule
Hybrid & 21.13 & 35.99 & 12.06 & 34.36 & 1.34 & 2.05 & 7.73 & 21.97 & 10.57 & 23.59 & 5.0 \\
HHR    & 20.31 & 35.57 & 19.28 & 54.25 & 8.04 & 14.20 & 8.66 & 25.49 & 14.07 & 32.38 & 5.0 \\
Dense  & \underline{31.55} & \textbf{52.21} & \underline{28.56} & \textbf{78.69} & \underline{11.03} & \underline{20.87} & \underline{10.82} & \underline{29.92} & \underline{20.49} & \underline{45.42} & 5.0 \\
\ours  & \textbf{55.79} & \underline{49.94} & \textbf{35.46} & \underline{53.31} & \textbf{33.70} & \textbf{36.80} & \textbf{40.74} & \textbf{53.95} & \textbf{41.42} & \textbf{48.50} & \textbf{3.2} \\
\bottomrule
\end{tabular}
}
\caption{Average Precision and Recall for the LegalBench-RAG dataset. The best performance values are highlighted in bold and the second-best are underlined.}
\label{table:legalbench}

\end{table*}

%% file: latex/appendix/case_study_failure.tex
\section{Failure Case Study}
\label{failure_case_study}

\paragraph{Retrieval Failure Case}
In this case, the model fails to retrieve all relevant documents across multiple fiscal years. The question explicitly asks for the total common dividends paid over the past three years (as of 2023), which requires retrieving information from multiple annual filings. However, the model only retrieves passages from the 2023 report. This indicates difficulty in reasoning over temporal scope and aggregating evidence when relevant information is distributed across documents.
\paragraph{Generation Failure Case}
In this case, the model retrieves a passage containing sufficient information to answer the question. However, the generated answer is incorrect due to faulty arithmetic reasoning. Specifically, the free cash flow should be computed as the difference between cash from operations and capital expenditures. Although both values are retrieved, the model produces an incorrect value, indicating limitations in table-grounded numerical reasoning and code execution.

\begin{table*}[!]
\footnotesize
\setlength{\fboxsep}{0pt}

\resizebox{\textwidth}{!}{%
\begin{tabular}{m{1.8cm} p{12.5cm}}
\toprule
\raggedright \textbf{Step} & \textbf{Content} \\
\midrule
\raggedright \textbf{Question} & 
\textit{How much common dividends did Bank of America pay in the last 3 years, as of 2023, in millions of US dollars?} \\
\midrule
\raggedright \textbf{Gold Evidences} & 
BAC\_2023\_10K (p.94), BAC\_2022\_10K (p.94), BAC\_2021\_10K (p.94) \\
\midrule
\raggedright \textbf{Retrieval} & 
\textbf{\textcolor{red}{\xmark} \;BAC\_2023\_10K (p.141):} NOTE 13 Shareholders’ Equity\newline Common Stock … \vspace{3pt} \\

& \textbf{\textcolor{red}{\xmark} \;BAC\_2023\_10K (p.150):} Defined Contribution Plans\newline The Corporation… \vspace{3pt} \\
\midrule
\raggedright \textbf{Generation} & 
\textbf{Ground Truth Answer}: 25,718 \quad \textbf{Generated Answer}: 2,560 \\
\bottomrule
\end{tabular}
}
\caption{Retrieval failure: model fails to retrieve documents from multiple years required to answer a multi-year aggregation question.}
\label{tab:retrieval-failure}
\end{table*}

\begin{table*}[!]
\footnotesize
\renewcommand{\arraystretch}{0.6}
\setlength{\fboxsep}{0pt}

\resizebox{\textwidth}{!}{%
\begin{tabular}{m{1.8cm} p{12.5cm}}
\toprule
\raggedright \textbf{Step} & \textbf{Content} \\
\midrule
\raggedright \textbf{Question} & 
\textit{According to the information provided in the statement of cash flows, what is the FY2020 free cash flow (FCF) for General Mills? Answer in USD millions.} \\
\midrule
\raggedright \textbf{Gold Evidences} & 
GIS\_2020\_10K (p.51) \\
\midrule
\raggedright \textbf{Retrieval} & 
\textbf{\textcolor{green}{\cmark} \;GIS\_2020\_10K (p.51):} … \$\textbf{3,676.2}… Capital expenditures \$\textbf{460.8} … \vspace{3pt} \\

& \textbf{\textcolor{red}{\xmark} \;GIS\_2020\_10K (p.17):} Adjusted diluted EPS of \$3.61 increased 12 percent on a constant-currency basis … \vspace{3pt} \\

& \textbf{\textcolor{red}{\xmark} \;GIS\_2020\_10K (p.36):} Free cash flow \$3,215.4 … Net cash provided by operating activities conversion … \\
\midrule
\raggedright \textbf{LLM Output} & 
\texttt{cash\_from\_operations = 3700 \# in USD millions\newline capex = 461 \# in USD millions} \\
\midrule
\raggedright \textbf{Generation} & 
\textbf{Ground Truth Answer}: 3,215 \quad \textbf{Generated Answer}: 3,239 \\
\bottomrule
\end{tabular}
}
\caption{Generation failure: despite correct retrieval, the model fails to compute free cash flow accurately from table values.}
\label{tab:generation-failure}
\end{table*}

%% file: tables/category.tex
\begin{tabular}{p{3cm} p{11cm}}
    \toprule
    \textbf{Category} & \textbf{Examples} \\
    \midrule
    Numeric (Table) & 
    \textbf{Question:} What percentage of the estimated purchase price for Hologic in 2007 is attributed to the net tangible assets? \\[0.5ex]
    & \textbf{Answer:} 3.8\% \\[0.5ex]
    & \textbf{Evidence:} HOLX\_2007\_10K (p.110) \\
    & The components and initial allocation of the purchase price consist of: \\ 
    & \fcolorbox{blue!50}{blue!10}{\strut Net tangible assets acquired as of September 18, 2007: \$2,800} \\ 
    & Developed technology and know how: \$12,300 \\ 
    & Customer relationship: \$17,000 \\ 
    & Trade name: \$2,800 \\ 
    & \ldots \\ 
    & \fcolorbox{blue!50}{blue!10}{\strut Estimated Purchase Price: \$73,200.} \\[1ex]
    \hline
    Numeric (Text) & 
    \textbf{Question:} During the years 2004 to 2006, what was the average impairment on construction in progress, expressed in millions, for American Tower Corporation as reported in their 2006 financial documents? \\[0.5ex]
    & \textbf{Answer:} 2.63 \\[0.5ex]
    & \textbf{Evidence:} AMT\_2006\_10K (p.106) \\
    & Construction-In-Progress Impairment Charges—For the years ended December 31, 2006, 2005 and 2004, the Company wrote-off approximately \fcolorbox{blue!50}{blue!10}{\strut \$1.0 million, \$2.3 million and \$4.6 million}, respectively, of construction-in-progress costs, primarily associated with sites that it no longer planned to build. \\[1ex]
    \hline
    Textual & 
    \textbf{Question:} What are the geographies that American Express primarily operates in as of 2022? \\[0.5ex]
    & \textbf{Answer:} United States, EMEA, APAC, and LACC \\[0.5ex]
    & \textbf{Evidence:} AXP\_2022\_10K (p.154) \\
    &Effective for the first quarter of 2022, we changed the way in which we allocate certain overhead expenses by geographic region. As a result, prior period pretax income (loss) from continuing operations by geography has been recast to conform to current period presentation there was no impact at a consolidated level. \\ 
    & (Millions) \fcolorbox{blue!50}{blue!10}{\strut United States EMEA APAC LACC} Other Unallocated Consolidated \\
    & 2022 \\
    & Total revenues net of interest expense $ 41,396 $ 4,871 $ 3,835 $ 2,917 $ (157) $ 52,862 \\
    \bottomrule
\end{tabular}
\caption{Examples of financial QA pairs categorized by question type. The highlighted segments in the evidence indicate the spans that are most relevant to the question.}